\documentclass[aps,prl,preprint,superscriptaddress,preprintnumbers,nofootinbib,tightenlines,floatfix]{revtex4-1}

\usepackage{graphicx} 
\usepackage{dcolumn}  
\usepackage{bm}       

\newcommand{\mevcc}{\mathrm{MeV}/c^2}

\newcommand{\mev}{\mathrm{MeV}}

\newcommand{\gevc}{\mathrm{GeV}/c}
\newcommand{\gev}{\mathrm{GeV}}

\newcommand{\ipb}{\mathrm{pb^{-1}}}
\newcommand{\pb}{\mathrm{pb}}

\newcommand{\epem}{e^+e^-}
\newcommand{\pipihc}{\pi^+\pi^-h_c}
\newcommand{\eepipihc}{\epem\to\pipihc}
\newcommand{\hcetac}{h_c\to\gamma\eta_c}
\newcommand{\pizhc}{\pi^0h_c}
\newcommand{\etahc}{\eta h_c}
\newcommand{\eeetahc}{\epem\to\etahc}
\newcommand{\xxx}{(\pi^0\pi^0/\pi^0/\eta)}
\newcommand{\xxxx}{X}
\newcommand{\xxxhc}{\xxx h_c}
\newcommand{\xxxxhc}{\xxxx h_c}
\newcommand{\eexxxxhc}{\epem\to\xxxx h_c}
\newcommand{\psipizhc}{\psi(2S)\to\pizhc}

\newcommand{\modei}{2(\pi^+\pi^-)}
\newcommand{\modeii}{2(\pi^+\pi^-)2\pi^0}
\newcommand{\modeiii}{3(\pi^+\pi^-)}
\newcommand{\modeiv}{K^{\pm}K_S^0\pi^{\mp}}
\newcommand{\modev}{K^{\pm}K_S^0\pi^{\mp}\pi^+\pi^-}
\newcommand{\modevi}{K^+K^-\pi^0}
\newcommand{\modevii}{K^+K^-\pi^+\pi^-}
\newcommand{\modeviii}{K^+K^-\pi^+\pi^-\pi^0}
\newcommand{\modeix}{K^+K^-2(\pi^+\pi^-)}
\newcommand{\modex}{2(K^+K^-)}
\newcommand{\modexi}{\eta\pi^+\pi^-}
\newcommand{\modexii}{\eta2(\pi^+\pi^-)}

\newcommand{\numpsi}{N^{\pi^0}_{\psi}}

\newcommand{\numex}{N^{X}_{E}}

\newcommand{\numpipi}{N^{\pi^+\pi^-}_{4170}}

\newcommand{\effratio}{R_{\epsilon}}
\newcommand{\numpsidecays}{N_{\psi}}
\newcommand{\lume}{{\cal L}_E}
\newcommand{\xxex}{\sigma^{X}_E}
\newcommand{\bpsi}{{\cal B}^{\pi^0}_{\psi}}

\newcommand{\ecm}{E_{CM}}
\newcommand{\chidof}{\chi_{4C}^2/\mathrm{d.o.f.}}

\newcommand{\resulta}{15.6\pm2.3\pm1.9\pm3.0~\pb}
\newcommand{\resultb}{131\pm15}
\newcommand{\resultc}{3523.86\pm0.48~\mevcc}
\newcommand{\resultd}{3523.86~\mevcc}
\newcommand{\resulte}{202\pm16}
\newcommand{\resultf}{3525.27\pm0.17~\mevcc}

\begin{document}

\preprint{CLNS 11/2073}  
\preprint{CLEO 11-1 }    

\title{\boldmath Observation of the $h_c(1P)$ using $e^+e^-$ collisions above $D\bar{D}$ threshold}



\author{T.~K.~Pedlar}
\affiliation{Luther College, Decorah, Iowa 52101, USA}
\author{D.~Cronin-Hennessy}
\author{J.~Hietala}
\affiliation{University of Minnesota, Minneapolis, Minnesota 55455, USA}
\author{S.~Dobbs}
\author{Z.~Metreveli}
\author{K.~K.~Seth}
\author{A.~Tomaradze}
\author{T.~Xiao}
\affiliation{Northwestern University, Evanston, Illinois 60208, USA}
\author{L.~Martin}
\author{A.~Powell}
\author{G.~Wilkinson}
\affiliation{University of Oxford, Oxford OX1 3RH, UK}
\author{H.~Mendez}
\affiliation{University of Puerto Rico, Mayaguez, Puerto Rico 00681}
\author{J.~Y.~Ge}
\author{D.~H.~Miller}
\author{I.~P.~J.~Shipsey}
\author{B.~Xin}
\affiliation{Purdue University, West Lafayette, Indiana 47907, USA}
\author{G.~S.~Adams}
\author{D.~Hu}
\author{B.~Moziak}
\author{J.~Napolitano}
\affiliation{Rensselaer Polytechnic Institute, Troy, New York 12180, USA}
\author{K.~M.~Ecklund}
\affiliation{Rice University, Houston, Texas 77005, USA}
\author{J.~Insler}
\author{H.~Muramatsu}
\author{C.~S.~Park}
\author{L.~J.~Pearson}
\author{E.~H.~Thorndike}
\affiliation{University of Rochester, Rochester, New York 14627, USA}
\author{S.~Ricciardi}
\affiliation{STFC Rutherford Appleton Laboratory, Chilton, Didcot, Oxfordshire, OX11 0QX, UK}
\author{C.~Thomas}
\affiliation{University of Oxford, Oxford OX1 3RH, UK}
\affiliation{STFC Rutherford Appleton Laboratory, Chilton, Didcot, Oxfordshire, OX11 0QX, UK}
\author{M.~Artuso}
\author{S.~Blusk}
\author{R.~Mountain}
\author{T.~Skwarnicki}
\author{S.~Stone}
\author{L.~M.~Zhang}
\affiliation{Syracuse University, Syracuse, New York 13244, USA}
\author{G.~Bonvicini}
\author{D.~Cinabro}
\author{A.~Lincoln}
\author{M.~J.~Smith}
\author{P.~Zhou}
\author{J.~Zhu}
\affiliation{Wayne State University, Detroit, Michigan 48202, USA}
\author{P.~Naik}
\author{J.~Rademacker}
\affiliation{University of Bristol, Bristol BS8 1TL, UK}
\author{D.~M.~Asner}
\altaffiliation[Now at: ]{Pacific Northwest National Laboratory, Richland, WA 99352}
\author{K.~W.~Edwards}
\author{K.~Randrianarivony}
\author{G.~Tatishvili}
\altaffiliation[Now at: ]{Pacific Northwest National Laboratory, Richland, WA 99352}
\affiliation{Carleton University, Ottawa, Ontario, Canada K1S 5B6}
\author{R.~A.~Briere}
\author{H.~Vogel}
\affiliation{Carnegie Mellon University, Pittsburgh, Pennsylvania 15213, USA}
\author{P.~U.~E.~Onyisi}
\author{J.~L.~Rosner}
\affiliation{University of Chicago, Chicago, Illinois 60637, USA}
\author{J.~P.~Alexander}
\author{D.~G.~Cassel}
\author{S.~Das}
\author{R.~Ehrlich}
\author{L.~Gibbons}
\author{S.~W.~Gray}
\author{D.~L.~Hartill}
\author{B.~K.~Heltsley}
\author{D.~L.~Kreinick}
\author{V.~E.~Kuznetsov}
\author{J.~R.~Patterson}
\author{D.~Peterson}
\author{D.~Riley}
\author{A.~Ryd}
\author{A.~J.~Sadoff}
\author{X.~Shi}
\author{W.~M.~Sun}
\affiliation{Cornell University, Ithaca, New York 14853, USA}
\author{J.~Yelton}
\affiliation{University of Florida, Gainesville, Florida 32611, USA}
\author{P.~Rubin}
\affiliation{George Mason University, Fairfax, Virginia 22030, USA}
\author{N.~Lowrey}
\author{S.~Mehrabyan}
\author{M.~Selen}
\author{J.~Wiss}
\affiliation{University of Illinois, Urbana-Champaign, Illinois 61801, USA}
\author{J.~Libby}
\affiliation{Indian Institute of Technology Madras, Chennai, Tamil Nadu 600036, India}
\author{M.~Kornicer}
\author{R.~E.~Mitchell}
\author{M.~R.~Shepherd}
\author{C.~M.~Tarbert}
\affiliation{Indiana University, Bloomington, Indiana 47405, USA }
\author{D.~Besson}
\affiliation{University of Kansas, Lawrence, Kansas 66045, USA}
\collaboration{CLEO Collaboration}
\noaffiliation


\date{June 21, 2011}

\begin{abstract} 
Using $586~\ipb$ of $\epem$ collision data at $\ecm = 4170~\mev$, produced at the CESR collider and collected with the CLEO-c detector, we observe the process $\eepipihc(1P)$. 
We measure its cross section to be $\resulta$, where the third error is due to the external uncertainty on the branching fraction of $\psipizhc(1P)$, which we use for normalization.  We also find evidence for $\eeetahc(1P)$ at $4170~\mev$ at the $3\sigma$ level, and see hints of a rise in the $\eepipihc(1P)$ cross section at $4260~\mev$.
\end{abstract}

\maketitle

In a previous Letter~\cite{Coan:2006rv}, the CLEO Collaboration investigated fifteen transitions to the $J/\psi$, $\psi(2S)$, and $\chi_{cJ}$ of charmonium states produced in $\epem$ collisions with $\ecm=3970-4260~\mev$.  The data were grouped into three energy bins ($3970-4060$, $4120-4200$, and $4260~\mev$) roughly corresponding to the $\psi(4040)$, $\psi(4160)$ and $Y(4260)$ regions.  Increases in the $\epem\to\pi^+\pi^-J/\psi$ and $\epem\to\pi^0\pi^0J/\psi$ cross sections at $\ecm=4260~\mev$ were attributed to $Y(4260)$ production~\cite{Aubert:2005rm}.  In this Letter, we extend those investigations to search for $\pi^+\pi^-$, $\pi^0\pi^0$, $\pi^0$, and $\eta$ transitions to the $h_c$ (where $h_c\equiv h_c(1P)$).  We use the same $60~\ipb$ of data with $\ecm=3970-4260~\mev$ (referred to as the ``scan data''~\cite{CroninHennessy:2008yi}) and the same energy binning, but we also now use $586~\ipb$ of data collected at $\ecm=4170~\mev$ (referred to as the ``4170 data'').  The 4170 data set is an order of magnitude larger than was available at that energy for the previous study.  The observation of transitions to the $h_c$ could provide insight into the perplexing nature of the charmonium states above $D\bar{D}$ threshold~\cite{Brambilla:2010cs}. It has also inspired new ways to search for and study bottomonium states, such as the $h_b$~\cite{:2011zp,Collaboration:2011ji}.

We search for the processes $\epem\to X h_c~(X\equiv\pi^+\pi^-,~\pi^0\pi^0,~\pi^0,~\eta)$  by reconstructing the $h_c$ through $\gamma\eta_c$ and the $\eta_c$ through: $\modei$, $\modeii$, $\modeiii$, $\modeiv$, $\modev$, $\modevi$, $\modevii$, $\modeviii$, $\modeix$, $\modex$, $\modexi$, and $\modexii$, the same twelve modes used in the CLEO measurement of ${\cal B}(J/\psi\to\gamma\eta_c)$~\cite{:2008fb}.  We also use a data sample of $24.5$~million $\psi(2S)$ decays to reconstruct the process $\psipizhc$ using the same method. To eliminate dependence on the branching fractions of the $\eta_c$, we take ratios of the cross sections~($\xxex$) for $\epem\to\xxxxhc$ at center-of-mass energy $E$ to the branching fraction~($\bpsi$) of $\psipizhc$.  We use $\bpsi = (8.4\pm1.3\pm1.0)\times10^{-4}$, measured by BESIII~\cite{Ablikim:2010rc}, to obtain $\xxex$.

We utilize symmetric $\epem$ collisions provided by the Cornell Electron Storage Ring~(CESR) with center-of-mass energies at the $\psi(2S)$ mass and in the range $3970-4260~\mev$.  The resulting final state particles ($K^{\pm}$, $\pi^{\pm}$, and $\gamma$) are detected by the CLEO-c detector~\cite{Kubota:1991ww}, which has a solid angle coverage of $93\%$.  The momenta of charged particles are measured by concentric drift chambers~\cite{Peterson:2002sk}, operating in a $1.0~\!\mathrm{T}$ magnetic field along the beam axis, with relative momentum resolutions of $\approx\!0.6\%$ at $p=1~\gevc$. To separate $K^\pm$ from $\pi^\pm$, two particle identification systems are used --  one based on ionization energy loss ($dE/dx$) in the drift chamber and the other a ring-imaging Cherenkov (RICH) detector~\cite{Artuso:2002ya}.  Photon energies are measured with a  cesium iodide calorimeter, which has relative energy resolutions of $2.2\%$ at $E_{\gamma}=1~\gev$ and $5\%$ at $100~\mev$.

We use standard track quality, particle identification, and calorimetry selection requirements~\cite{:2008fb} to reconstruct the exclusive processes $\epem\to\xxxxhc$ and $\psipizhc$ with $\hcetac$.  The $\eta$'s from the $\eta_c$ are reconstructed in both their $\gamma\gamma$ and $\pi^+\pi^-\pi^0$ decay modes, but the transition $\eta$ from $\eeetahc$ is only reconstructed in its $\gamma\gamma$ mode (due to large combinatoric backgrounds and small efficiencies for the $\pi^+\pi^-\pi^0$ mode).  For $\pi^0$ and $\eta$ decays to $\gamma\gamma$, the mass of the pair of daughter photons is required to be within $3\sigma$ of the nominal mass and is subsequently constrained to that mass.   To reconstruct $\eta\rightarrow\pi^+\pi^-\pi^0$, the three pions must have an invariant mass within $30~\mevcc$ of the nominal $\eta$ mass.  The $K_S^0$ candidates are selected from pairs of oppositely charged and vertex-constrained tracks (assumed to be pions) with invariant mass within $15~\mevcc$ of the $K_S^0$ mass.  In addition, we require that the photon from $\hcetac$ cannot be paired with any other shower in the event to form a diphoton mass within $3\sigma$ of the $\pi^0$ mass.  A four-constraint kinematic fit of all identified particles to the initial $\epem$ four-momentum is then performed and the resulting fit quality is required to satisfy $\chidof<5$.  This procedure sharpens the measured momenta in signal events and reduces backgrounds with missing or extra particles.  For each decay mode of the $\eta_c$, the candidate with the best fit quality is accepted. The selection criteria for $\psipizhc$ are identical to that for $\epem\to\xxxxhc$, except for an additional requirement suppressing $\psi(2S)\to\pi^+\pi^-J/\psi$ by the exclusion of any event with a $\pi^+\pi^-$ pair with a recoil mass within $15~\mevcc$ of $M(J/\psi)$.

\begin{figure}
\includegraphics[width=0.5\columnwidth]{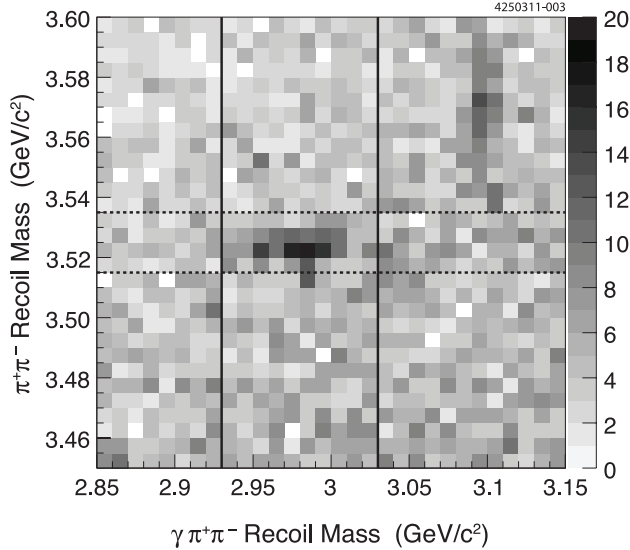}
\caption{\label{fig:HCETAC}The recoil mass of the $\pi^+\pi^-$ system {\it versus} the recoil mass of the $\gamma\pi^+\pi^-$ system for candidate $\eepipihc; \hcetac$ events at $\ecm=4170~\mev$.  The signal appears at the intersection of the $h_c$ and $\eta_c$ masses.  The vertical lines indicate the region used to select the $\eta_c$.  The horizontal lines mark $\pm10~\mevcc$ around the $h_c$ mass.}
\end{figure}

We select the $\eta_c$ by requiring the recoil mass of the $\gamma X$ system be between $2930$ and $3030~\mevcc$.  We then search for the $h_c$ in the recoil mass distribution of the $X$ system.  Figure~\ref{fig:HCETAC} shows a histogram of the $\pi^+\pi^-$ and $\gamma\pi^+\pi^-$ recoil masses for the process $\eepipihc$ at $\ecm=4170~\mev$.  A clear accumulation of events can be seen near the intersection of the $h_c$ and $\eta_c$ masses, which marks the signal.  Background from the initial state radiation process $\epem\to\gamma\psi(2S); \psi(2S)\to\pi^+\pi^-J/\psi$ appears as a vertical band at the $J/\psi$ mass and is well-separated from the signal.  Other backgrounds, studied with dedicated background Monte Carlo simulations, are smooth and are due to the light-quark continuum ($\epem\to q\bar{q}$) or $D\bar{D}$ production, simulated with previously measured cross sections~\cite{CroninHennessy:2008yi}.

\begin{figure}
\includegraphics[width=0.7\columnwidth]{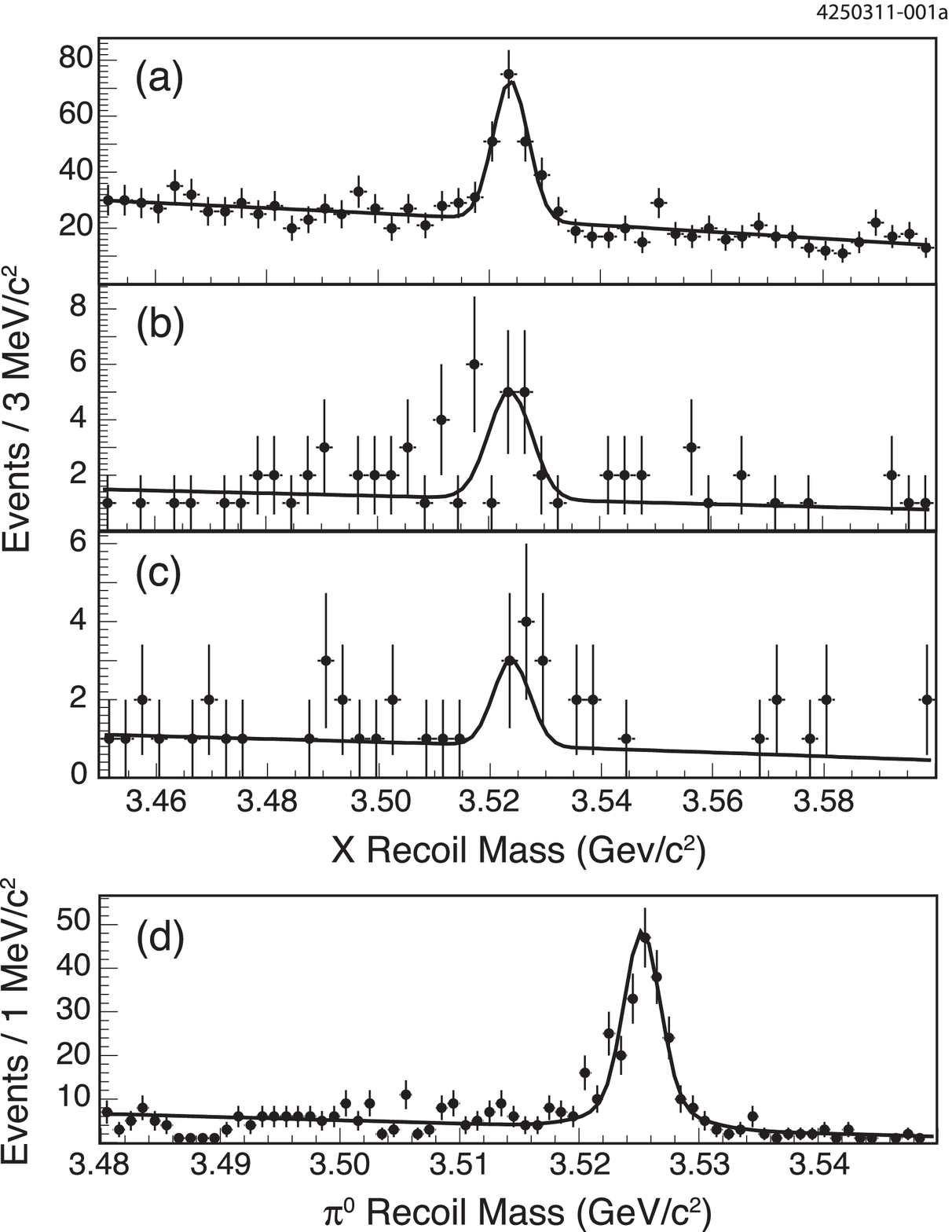}
\caption{\label{fig:N}Fits to determine the yields of $h_c$ events from (a)~$\eepipihc$ at $\ecm=4170~\mev$; (b)~$\eeetahc$ at $\ecm=4170~\mev$; (c)~$\eepipihc$ at $\ecm=4260~\mev$;  and (d)~the normalizing mode, $\psipizhc$.}
\end{figure}

The yield of $\eepipihc$ events at $\ecm = 4170~\mev$ is determined by fitting the $\pi^+\pi^-$ recoil mass distribution, after selecting the $\eta_c$, with two components.  The signal shape is described by a double Gaussian with floating mass and normalization, but with widths fixed by signal Monte Carlo.  The background shape is a freely floating first-order polynomial.  The resulting fit is shown in Fig.~\ref{fig:N}(a).  We find $\resultb$ signal events with a significance of more than $10\sigma$.  The significance, here and in subsequent fits, is calculated from log-likelihood differences between fits with and without a signal component.  The resulting mass from the fit is $\resultc$ (statistical errors only), which is $1.5~\mevcc$ lower than the PDG~2010 value of $3525.42\pm0.29~\mevcc$~\cite{pdg}.  This discrepancy, however, is less than the uncertainty of the initial $\epem$ energy~($\approx2~\mev$) used in the kinematic fit, which directly affects the measured dipion recoil mass.  

\begin{turnpage}
\begin{table*}
\caption{\label{tab:results} Yields~($\numex$), significances, relative fitting and shape systematic errors, efficiency ratios~($\effratio$), normalized cross sections~($\xxex/\bpsi$), and cross sections~($\xxex$) for each reaction $\epem\to Xh_c$. The third error on $\xxex$ is from $\bpsi$~\cite{Ablikim:2010rc}.}
\begin{ruledtabular}
\begin{tabular}{lcccccccc}
X & $\ecm$ & $\numex$ & Sig. & Fitting & Shape & $\effratio$ & $\xxex/\bpsi$ & $\xxex$ \\
& (MeV) & (Events) & ($\sigma$) & Syst. (\%) & Syst. (\%) & & (nb) & (pb) \\
\hline
$\pi^{0}$  &  3686 ($\psi(2S)$)  &  $202\pm16$  &  $>10$  &  4.8  &  3.9  &  --  &  --  &  --  \\
$\pi^{+}\pi^{-}$  &  4170  &  $131\pm15$  &  $>10$  &  1.7  &  7.1  &  $1.46\pm0.04$  &  $18.5\pm2.7\pm2.2$  &  $15.6\pm2.3\pm1.9\pm3.0$  \\
$\pi^{0}\pi^{0}$  &  4170  &  $7.4\pm8.0$  &  1.0  &  23  &  27  &  $0.43\pm0.02$  &  $3.6\pm3.9\pm1.4$  &  $3.0\pm3.3\pm1.1\pm0.6$  \\
$\pi^{0}$  &  4170  &  $-5\pm11$  &  --  &  47  &  77  &  $1.12\pm0.03$  &  $-0.9\pm2.1\pm0.8$  &  $-0.7\pm1.8\pm0.7\pm0.1$  \\
$\eta $  &  4170  &  $12.6\pm4.5$  &  3.8  &  13  &  11  &  $0.47\pm0.01$  &  $5.6\pm2.1\pm1.1$  &  $4.7\pm1.7\pm1.0\pm0.9$  \\
$\pi^{+}\pi^{-}$  &  3970--4060  &  $0.3\pm2.1$  &  0.1  &  400  &  360  &  $1.30\pm0.04$  &  $1.2\pm9.5\pm6.4$  &  $1.0\pm8.0\pm5.4\pm0.2$  \\
$\pi^{+}\pi^{-}$  &  4120--4200  &  $4.4\pm3.1$  &  1.7  &  52  &  27  &  $1.46\pm0.04$  &  $13.9\pm9.9\pm8.2$  &  $11.7\pm8.3\pm6.9\pm2.3$  \\
$\pi^{+}\pi^{-}$  &  4260  &  $6.0\pm3.1$  &  2.6  &  4.9  &  17  &  $1.49\pm0.04$  &  $38\pm20\pm8$  &  $32\pm17\pm6\pm6$  \\
\end{tabular}
\end{ruledtabular}
\end{table*}
\end{turnpage}

Fits to $\epem\to\xxxhc$ at $\ecm=4170~\mev$ and fits to $\eepipihc$ with $\ecm=3970-4260~\mev$ follow the same procedure except that, due to lower statistics, the mass is fixed to the value obtained previously, $\resultd$.  The resulting yields and significances are listed in Table~\ref{tab:results}.  We find $>\!3\sigma$ evidence for $\eeetahc$ at $4170~\mev$ (Fig.~\ref{fig:N}(b)) and hints of a signal ($2.6\sigma$) for $\eepipihc$ at $4260~\mev$ (Fig.~\ref{fig:N}(c)).

The normalizing mode, $\psipizhc$, is also fit using the same method and with a floating mass (Fig.~\ref{fig:N}(d)).  The yield is measured to be $\resulte$ events.  The resulting mass is $\resultf$ (statistical errors only), consistent with, and highly correlated to, a previous measurement by CLEO using a similar method~\cite{cleochc}.

We calculate the ratios of the cross sections of $\epem\to\xxxxhc$ at energy $E$ ($\xxex$) to the branching fraction of $\psipizhc$ ($\bpsi$) using:
\begin{equation}
\label{eq:method}
\frac{\xxex}{\bpsi} = \frac{\numpsidecays}{\lume}\frac{\numex}{\numpsi\effratio},
\end{equation}
where $\numpsidecays$ is the number of $\psi(2S)$ decays, $\lume$ is the luminosity at energy $E$, $\numex$ and $\numpsi$ are measured yields, and $\effratio$ is a ratio of selection efficiencies: that of $\eexxxxhc$ to that of $\psipizhc$.  Since the ratio of efficiencies for each $\eta_c$ decay mode is not perfectly constant (with $10\%-20\%$ variations), we weight the individual efficiency ratios by the number of $\psipizhc$ events we observe in each $\eta_c$ decay mode, which we obtain through the fitting procedure described above.  The errors on the efficiency ratios include errors due to Monte Carlo statistics and errors on these individual yields.

Previously determined systematic errors are used for $\numpsidecays$~(2\%)~\cite{:2008kb} and $\lume$~(1\%)~\cite{:2007zt}.  Most systematic errors on individual track and photon reconstruction efficiencies cancel in the ratio of efficiencies, $\effratio$.  However, for the transition particles, the $\xxxx$ in the numerator and the $\pi^0$ in the denominator, a 1\% relative error is assigned for each track and a 2\% error for each photon.  A conservative 5\% systematic error is included for our determination of $\effratio$, which relies upon signal Monte Carlo distributed according to phase space.  This systematic error is estimated by using extreme variations of the $\eta_c$ substructure -- for example, by replacing $2(K^+K^-)$ by $\phi(1020)\phi(1020)$.

Systematic errors in $\numex$ and $\numpsi$ due to the fitting procedure are evaluated by varying the order of the background polynomials, varying the fit ranges, and varying the bin sizes.  Based on Monte Carlo studies, we also use background shapes determined by $\chidof$ sidebands ($10<\chidof<35$). For $\numpsi$, we alternatively use an ARGUS distribution~\cite{Albrecht:1994tb} for the background.

Systematic errors due to signal shapes are evaluated by varying the signal mass and width. The largest deviations occur when the signal widths are allowed to float.  This variation determines the shape systematic error on $\numpsi$ and $\numpipi$.  For other $\numex$, where the statistics are lower, the width variation is performed by scaling the width by the deviation observed between data and signal Monte Carlo in the fit for $\numpipi$, which is $\approx20\%$.  Variations of the signal mass produce smaller deviations.

\begin{figure}
\includegraphics[width=0.7\columnwidth]{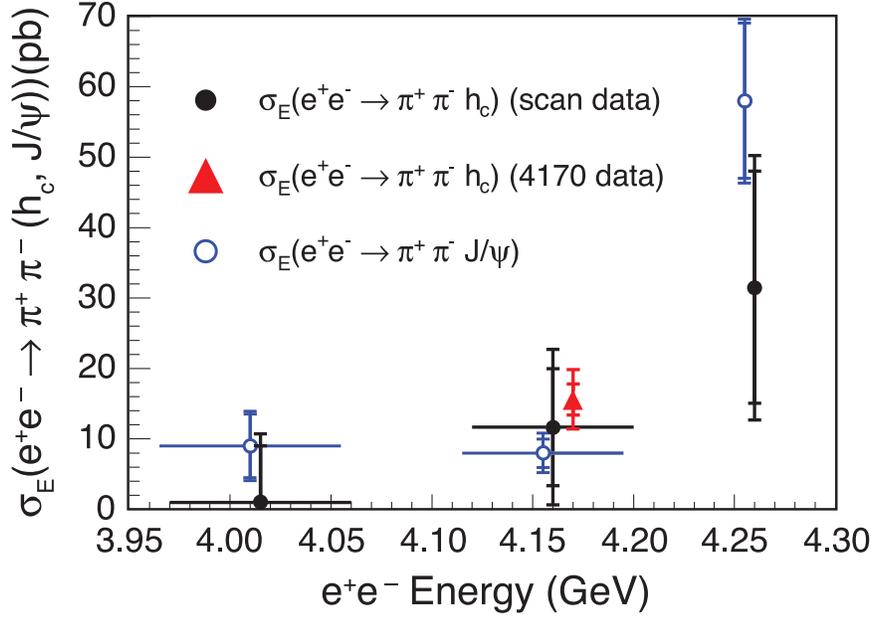}
\caption{\label{fig:SCAN}Cross sections as a function of center-of-mass energy.  The triangle shows the cross section for $\eepipihc$ at $\ecm=4170~\mev$; the closed circles are for the same process at other center-of-mass energies.  For reference, the $\epem\to\pi^+\pi^-J/\psi$ cross section~\cite{Coan:2006rv} is indicated by open circles.  The inner error bars are the statistical errors; the outer error bars are the quadratic sum of the statistical and systematic errors.}
\end{figure}

\begin{figure}[h!]
\includegraphics[width=1.0\columnwidth]{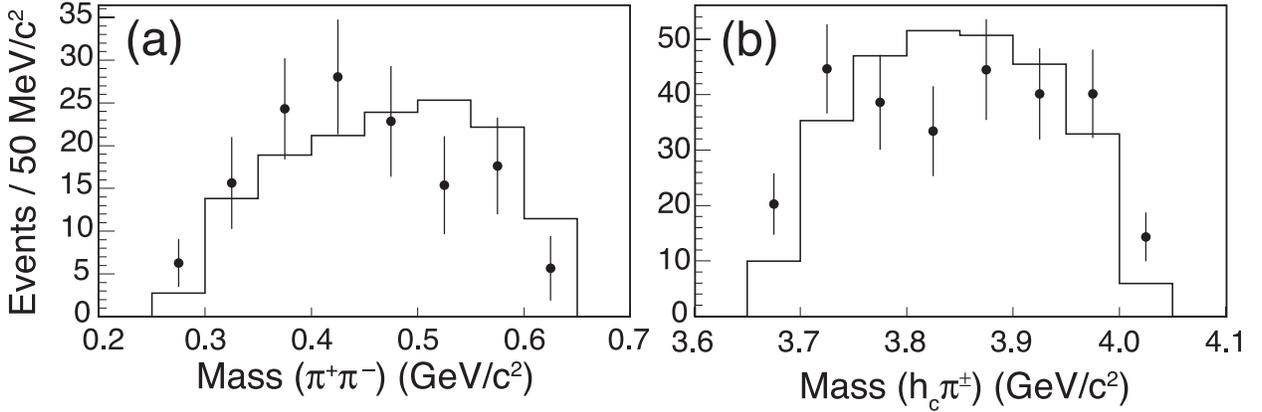}
\caption{\label{fig:DALITZ}The (a)~$\pi^+\pi^-$ and (b)~$h_c\pi^{\pm}$ mass distributions from $\eepipihc$ at $\ecm=4170~\mev$.  The points are obtained by fitting for the $h_c$ yields in bins of $\pi^+\pi^-$ or $\pi^{\pm}h_c$ mass. The histogram is signal~MC, generated according to phase space and scaled by the total $h_c$ yield.}
\end{figure}

The final numbers are listed in Table~\ref{tab:results}.  The $\pipihc$ cross sections as a function of center-of-mass energy are summarized in Fig.~\ref{fig:SCAN}.  Notice that the $\pipihc$ cross sections are of a comparable size to those of $\pi^+\pi^-J/\psi$.  There is also a suggestive rise in the cross section at $4260~\mev$, which could be an indication of $Y(4260)$ production, but will require further data to be definitive.

Projections of the $\pipihc$ Dalitz plot at $\ecm=4170~\mev$ are shown in Fig.~\ref{fig:DALITZ} and are compared to phase space Monte Carlo.  To separate signal from background, the number of signal $\pipihc$ events in each bin is determined by the fitting procedure described above.  The efficiency is relatively uniform across the Dalitz plot.  More data would be required to investigate any possible discrepancies of the data with phase space.

Assuming the $E_{CM} = 3970-4060~\mev$ and $E_{CM} = 4170~\mev$ data correspond to $\psi(4040)$ and $\psi(4160)$ production, respectively, we convert cross sections to upper limits on branching fractions using the same conversion factors listed in a previous CLEO analysis of this region~\cite{Coan:2006rv}.  The results are listed in Table~\ref{tab:b}. Assuming the $4260~\mev$ point is purely due to $Y(4260)$ production, we set a limit on its branching fraction to $\pi^+\pi^-h_c$ relative to $\pi^+\pi^-J/\psi$ of $<1.0$ at 90\% confidence level~(C.L.).

In summary, we observe the process $\eepipihc$ at $\ecm=4170~\mev$ and find its cross section to be comparable to the corresponding cross section for $J/\psi$ production.  This has already resulted in new methods to search for and study the $h_b$ using $\epem$ collisions above $B\bar{B}$ threshold~\cite{Collaboration:2011ji}.  We also see hints of a rise in the $\pipihc$ cross section at $\ecm=4260~\mev$.  Further data will be required, however, to determine if this rise can be attributed to the $Y(4260)$.

\begin{table}
\caption{\label{tab:b} Upper limits (at 90\% C.L.) on branching fractions for the $\psi(4040)$ and $\psi(4160)$ to $Xh_c$.}
\begin{ruledtabular}
\begin{tabular}{lcc}
X & ${\cal B}(\psi(4040)\to Xh_c)$ & ${\cal B}(\psi(4160)\to Xh_c)$  \\
& $(\times10^{-3})$ & $(\times10^{-3})$ \\
\hline
$\pi^+\pi^-$ & $<3$ & $<5$ \\
$\pi^0\pi^0$ & --   & $<2$ \\
$\pi^0$      & --   & $<0.4$ \\
$\eta$       & --   & $<2$ \\
\end{tabular}
\end{ruledtabular}
\end{table}

\begin{acknowledgments} 
We gratefully acknowledge the effort of the CESR staff 
in providing us with excellent luminosity and running conditions. 
This work was supported by the 
A.P.~Sloan Foundation, 
the National Science Foundation, 
the U.S. Department of Energy, 
the Natural Sciences and Engineering Research Council of Canada, and 
the U.K. Science and Technology Facilities Council. 
\end{acknowledgments} 

\bibliography{bib}

\begin{thebibliography}{16}%
\makeatletter
\providecommand \@ifxundefined [1]{%
 \@ifx{#1\undefined}
}%
\providecommand \@ifnum [1]{%
 \ifnum #1\expandafter \@firstoftwo
 \else \expandafter \@secondoftwo
 \fi
}%
\providecommand \@ifx [1]{%
 \ifx #1\expandafter \@firstoftwo
 \else \expandafter \@secondoftwo
 \fi
}%
\providecommand \natexlab [1]{#1}%
\providecommand \enquote  [1]{``#1''}%
\providecommand \bibnamefont  [1]{#1}%
\providecommand \bibfnamefont [1]{#1}%
\providecommand \citenamefont [1]{#1}%
\providecommand \href@noop [0]{\@secondoftwo}%
\providecommand \href [0]{\begingroup \@sanitize@url \@href}%
\providecommand \@href[1]{\@@startlink{#1}\@@href}%
\providecommand \@@href[1]{\endgroup#1\@@endlink}%
\providecommand \@sanitize@url [0]{\catcode `\\12\catcode `\$12\catcode
  `\&12\catcode `\#12\catcode `\^12\catcode `\_12\catcode `\%12\relax}%
\providecommand \@@startlink[1]{}%
\providecommand \@@endlink[0]{}%
\providecommand \url  [0]{\begingroup\@sanitize@url \@url }%
\providecommand \@url [1]{\endgroup\@href {#1}{\urlprefix }}%
\providecommand \urlprefix  [0]{URL }%
\providecommand \Eprint [0]{\href }%
\providecommand \doibase [0]{http://dx.doi.org/}%
\providecommand \selectlanguage [0]{\@gobble}%
\providecommand \bibinfo  [0]{\@secondoftwo}%
\providecommand \bibfield  [0]{\@secondoftwo}%
\providecommand \translation [1]{[#1]}%
\providecommand \BibitemOpen [0]{}%
\providecommand \bibitemStop [0]{}%
\providecommand \bibitemNoStop [0]{.\EOS\space}%
\providecommand \EOS [0]{\spacefactor3000\relax}%
\providecommand \BibitemShut  [1]{\csname bibitem#1\endcsname}%
\let\auto@bib@innerbib\@empty
\bibitem [{\citenamefont {Coan}\ \emph {et~al.}(2006)\citenamefont {Coan} \emph
  {et~al.}}]{Coan:2006rv}%
  \BibitemOpen
  \bibfield  {author} {\bibinfo {author} {\bibfnamefont {T.~E.}\ \bibnamefont
  {Coan}} \emph {et~al.} (\bibinfo {collaboration} {CLEO Collaboration}),\
  }\href {\doibase 10.1103/PhysRevLett.96.162003} {\bibfield  {journal}
  {\bibinfo  {journal} {Phys. Rev. Lett.}\ }\textbf {\bibinfo {volume} {96}},\
  \bibinfo {pages} {162003} (\bibinfo {year} {2006})},\ \Eprint
  {http://arxiv.org/abs/hep-ex/0602034} {arXiv:hep-ex/0602034} \BibitemShut
  {NoStop}%
\bibitem [{\citenamefont {Aubert}\ \emph {et~al.}(2005)\citenamefont {Aubert}
  \emph {et~al.}}]{Aubert:2005rm}%
  \BibitemOpen
  \bibfield  {author} {\bibinfo {author} {\bibfnamefont {B.}~\bibnamefont
  {Aubert}} \emph {et~al.} (\bibinfo {collaboration} {BABAR Collaboration}),\
  }\href {\doibase 10.1103/PhysRevLett.95.142001} {\bibfield  {journal}
  {\bibinfo  {journal} {Phys. Rev. Lett.}\ }\textbf {\bibinfo {volume} {95}},\
  \bibinfo {pages} {142001} (\bibinfo {year} {2005})},\ \Eprint
  {http://arxiv.org/abs/hep-ex/0506081} {arXiv:hep-ex/0506081} \BibitemShut
  {NoStop}%
\bibitem [{\citenamefont {Cronin-Hennessy}\ \emph {et~al.}(2009)\citenamefont
  {Cronin-Hennessy} \emph {et~al.}}]{CroninHennessy:2008yi}%
  \BibitemOpen
  \bibfield  {author} {\bibinfo {author} {\bibfnamefont {D.}~\bibnamefont
  {Cronin-Hennessy}} \emph {et~al.} (\bibinfo {collaboration} {CLEO
  Collaboration}),\ }\href {\doibase 10.1103/PhysRevD.80.072001} {\bibfield
  {journal} {\bibinfo  {journal} {Phys. Rev. D}\ }\textbf {\bibinfo {volume}
  {80}},\ \bibinfo {pages} {072001} (\bibinfo {year} {2009})},\ \Eprint
  {http://arxiv.org/abs/0801.3418} {arXiv:0801.3418 [hep-ex]} \BibitemShut
  {NoStop}%
\bibitem [{\citenamefont {Brambilla}\ \emph {et~al.}(2011)\citenamefont
  {Brambilla} \emph {et~al.}}]{Brambilla:2010cs}%
  \BibitemOpen
  \bibfield  {author} {\bibinfo {author} {\bibfnamefont {N.}~\bibnamefont
  {Brambilla}} \emph {et~al.},\ }\href {\doibase
  10.1140/epjc/s10052-010-1534-9} {\bibfield  {journal} {\bibinfo  {journal}
  {Eur. Phys. J. C}\ }\textbf {\bibinfo {volume} {71}},\ \bibinfo {pages}
  {1534} (\bibinfo {year} {2011})},\ \Eprint {http://arxiv.org/abs/1010.5827}
  {arXiv:1010.5827 [hep-ph]} \BibitemShut {NoStop}%
\bibitem [{\citenamefont {Lees}\ \emph {et~al.}(2011)\citenamefont {Lees} \emph
  {et~al.}}]{:2011zp}%
  \BibitemOpen
  \bibfield  {author} {\bibinfo {author} {\bibfnamefont {J.~P.}\ \bibnamefont
  {Lees}} \emph {et~al.} (\bibinfo {collaboration} {BABAR Collaboration}),\
  }\href@noop {} {\  (\bibinfo {year} {2011})},\ \Eprint
  {http://arxiv.org/abs/1102.4565} {arXiv:1102.4565 [hep-ex]} \BibitemShut
  {NoStop}%
\bibitem [{\citenamefont {Adachi}\ \emph {et~al.}(2011)\citenamefont {Adachi}
  \emph {et~al.}}]{Collaboration:2011ji}%
  \BibitemOpen
  \bibfield  {author} {\bibinfo {author} {\bibfnamefont {I.}~\bibnamefont
  {Adachi}} \emph {et~al.} (\bibinfo {collaboration} {Belle Collaboration}),\
  }\href@noop {} {\  (\bibinfo {year} {2011})},\ \Eprint
  {http://arxiv.org/abs/1103.3419} {arXiv:1103.3419 [hep-ex]} \BibitemShut
  {NoStop}%
\bibitem [{\citenamefont {Mitchell}\ \emph {et~al.}(2009)\citenamefont
  {Mitchell} \emph {et~al.}}]{:2008fb}%
  \BibitemOpen
  \bibfield  {author} {\bibinfo {author} {\bibfnamefont {R.~E.}\ \bibnamefont
  {Mitchell}} \emph {et~al.} (\bibinfo {collaboration} {CLEO Collaboration}),\
  }\href {\doibase 10.1103/PhysRevLett.102.011801} {\bibfield  {journal}
  {\bibinfo  {journal} {Phys. Rev. Lett.}\ }\textbf {\bibinfo {volume} {102}},\
  \bibinfo {pages} {011801} (\bibinfo {year} {2009})},\ \Eprint
  {http://arxiv.org/abs/0805.0252} {arXiv:0805.0252 [hep-ex]} \BibitemShut
  {NoStop}%
\bibitem [{\citenamefont {Ablikim}\ \emph {et~al.}(2010)\citenamefont {Ablikim}
  \emph {et~al.}}]{Ablikim:2010rc}%
  \BibitemOpen
  \bibfield  {author} {\bibinfo {author} {\bibfnamefont {M.}~\bibnamefont
  {Ablikim}} \emph {et~al.} (\bibinfo {collaboration} {BESIII Collaboration}),\
  }\href {\doibase 10.1103/PhysRevLett.104.132002} {\bibfield  {journal}
  {\bibinfo  {journal} {Phys. Rev. Lett.}\ }\textbf {\bibinfo {volume} {104}},\
  \bibinfo {pages} {132002} (\bibinfo {year} {2010})},\ \Eprint
  {http://arxiv.org/abs/1002.0501} {arXiv:1002.0501 [hep-ex]} \BibitemShut
  {NoStop}%
\bibitem [{\citenamefont {Kubota}\ \emph {et~al.}(1992)\citenamefont {Kubota}
  \emph {et~al.}}]{Kubota:1991ww}%
  \BibitemOpen
  \bibfield  {author} {\bibinfo {author} {\bibfnamefont {Y.}~\bibnamefont
  {Kubota}} \emph {et~al.} (\bibinfo {collaboration} {CLEO Collaboration}),\
  }\href {\doibase 10.1016/0168-9002(92)90770-5} {\bibfield  {journal}
  {\bibinfo  {journal} {Nucl. Instrum. Meth. A}\ }\textbf {\bibinfo {volume}
  {320}},\ \bibinfo {pages} {66} (\bibinfo {year} {1992})}\BibitemShut
  {NoStop}%
\bibitem [{\citenamefont {Peterson}\ \emph {et~al.}(2002)\citenamefont
  {Peterson} \emph {et~al.}}]{Peterson:2002sk}%
  \BibitemOpen
  \bibfield  {author} {\bibinfo {author} {\bibfnamefont {D.}~\bibnamefont
  {Peterson}} \emph {et~al.},\ }\href {\doibase 10.1016/S0168-9002(01)01737-5}
  {\bibfield  {journal} {\bibinfo  {journal} {Nucl. Instrum. Meth. A}\ }\textbf
  {\bibinfo {volume} {478}},\ \bibinfo {pages} {142} (\bibinfo {year}
  {2002})}\BibitemShut {NoStop}%
\bibitem [{\citenamefont {Artuso}\ \emph {et~al.}(2003)\citenamefont {Artuso}
  \emph {et~al.}}]{Artuso:2002ya}%
  \BibitemOpen
  \bibfield  {author} {\bibinfo {author} {\bibfnamefont {M.}~\bibnamefont
  {Artuso}} \emph {et~al.},\ }\href {\doibase 10.1016/S0168-9002(02)02162-9}
  {\bibfield  {journal} {\bibinfo  {journal} {Nucl. Instrum. Meth. A}\ }\textbf
  {\bibinfo {volume} {502}},\ \bibinfo {pages} {91} (\bibinfo {year} {2003})},\
  \Eprint {http://arxiv.org/abs/hep-ex/0209009} {arXiv:hep-ex/0209009}
  \BibitemShut {NoStop}%
\bibitem [{\citenamefont {Nakamura}\ \emph {et~al.}(2010)\citenamefont
  {Nakamura} \emph {et~al.}}]{pdg}%
  \BibitemOpen
  \bibfield  {author} {\bibinfo {author} {\bibfnamefont {K.}~\bibnamefont
  {Nakamura}} \emph {et~al.} (\bibinfo {collaboration} {Particle Data Group}),\
  }\href {\doibase 10.1088/0954-3899/37/7A/075021} {\bibfield  {journal}
  {\bibinfo  {journal} {J. Phys. G}\ }\textbf {\bibinfo {volume} {37}},\
  \bibinfo {pages} {075021} (\bibinfo {year} {2010})}\BibitemShut {NoStop}%
\bibitem [{\citenamefont {Dobbs}\ \emph {et~al.}(2008)\citenamefont {Dobbs}
  \emph {et~al.}}]{cleochc}%
  \BibitemOpen
  \bibfield  {author} {\bibinfo {author} {\bibfnamefont {S.}~\bibnamefont
  {Dobbs}} \emph {et~al.} (\bibinfo {collaboration} {CLEO Collaboration}),\
  }\href {\doibase 10.1103/PhysRevLett.101.182003} {\bibfield  {journal}
  {\bibinfo  {journal} {Phys. Rev. Lett.}\ }\textbf {\bibinfo {volume} {101}},\
  \bibinfo {pages} {182003} (\bibinfo {year} {2008})},\ \Eprint
  {http://arxiv.org/abs/0805.4599} {arXiv:0805.4599 [hep-ex]} \BibitemShut
  {NoStop}%
\bibitem [{\citenamefont {Mendez}\ \emph {et~al.}(2008)\citenamefont {Mendez}
  \emph {et~al.}}]{:2008kb}%
  \BibitemOpen
  \bibfield  {author} {\bibinfo {author} {\bibfnamefont {H.}~\bibnamefont
  {Mendez}} \emph {et~al.} (\bibinfo {collaboration} {CLEO Collaboration}),\
  }\href {\doibase 10.1103/PhysRevD.78.011102} {\bibfield  {journal} {\bibinfo
  {journal} {Phys. Rev. D}\ }\textbf {\bibinfo {volume} {78}},\ \bibinfo
  {pages} {011102} (\bibinfo {year} {2008})},\ \Eprint
  {http://arxiv.org/abs/0804.4432} {arXiv:0804.4432 [hep-ex]} \BibitemShut
  {NoStop}%
\bibitem [{\citenamefont {Dobbs}\ \emph {et~al.}(2007)\citenamefont {Dobbs}
  \emph {et~al.}}]{:2007zt}%
  \BibitemOpen
  \bibfield  {author} {\bibinfo {author} {\bibfnamefont {S.}~\bibnamefont
  {Dobbs}} \emph {et~al.} (\bibinfo {collaboration} {CLEO Collaboration}),\
  }\href {\doibase 10.1103/PhysRevD.76.112001} {\bibfield  {journal} {\bibinfo
  {journal} {Phys. Rev. D}\ }\textbf {\bibinfo {volume} {76}},\ \bibinfo
  {pages} {112001} (\bibinfo {year} {2007})},\ \Eprint
  {http://arxiv.org/abs/0709.3783} {arXiv:0709.3783 [hep-ex]} \BibitemShut
  {NoStop}%
\bibitem [{\citenamefont {Albrecht}\ \emph {et~al.}(1994)\citenamefont
  {Albrecht} \emph {et~al.}}]{Albrecht:1994tb}%
  \BibitemOpen
  \bibfield  {author} {\bibinfo {author} {\bibfnamefont {H.}~\bibnamefont
  {Albrecht}} \emph {et~al.} (\bibinfo {collaboration} {ARGUS Collaboration}),\
  }\href {\doibase 10.1016/0370-2693(94)01302-0} {\bibfield  {journal}
  {\bibinfo  {journal} {Phys. Lett. B}\ }\textbf {\bibinfo {volume} {340}},\
  \bibinfo {pages} {217} (\bibinfo {year} {1994})}\BibitemShut {NoStop}%
\end{thebibliography}%





\end{document}